\documentstyle[aps,pra,epsf,floats]{revtex}
\def\figuresize{\ifpreprintsty 12cm \else 8cm \fi}

\begin{document}

\ifpreprintsty \else
\twocolumn[\hsize\textwidth\columnwidth\hsize\csname@twocolumnfalse%
\endcsname \fi

\draft
\title{In an Ising model with spin-exchange dynamics damage always spreads}

\author{Thomas Vojta}
\address{Institut f\"ur Physik, Technische Universit\"at, D-09107 Chemnitz, Germany}
\date{version June 2, printed \today}
\maketitle

\begin{abstract}
We investigate the spreading of damage in Ising models with Kawasaki spin-exchange dynamics
which conserves the magnetization.  We first modify a recent master equation approach 
to account for dynamic rules involving more than a single site. We then derive an 
effective-field theory for damage spreading in Ising models with Kawasaki spin-exchange dynamics 
and solve it for a two-dimensional model on a honeycomb lattice. In contrast to the cases of 
Glauber or heat-bath dynamics, we find that the damage always spreads and never
heals.  In the long-time limit the average Hamming distance approaches that of two uncorrelated 
systems. These results are verified by Monte-Carlo simulations.
\end{abstract}

\ifpreprintsty \else
] \fi              


\section{Introduction}

Damage spreading (DS) investigates how a small perturbation in a cooperative system 
changes during the time evolution \cite{kauffman,creutz,derrida1,stanley} (for a short review see, 
e.g., Ref. \cite{jan}).
In order to study  DS  two replicas of the system are considered which evolve stochastically 
under the same noise realization (i.e. the same random numbers are used in a  
Monte-Carlo procedure).  The difference in the microscopic configurations of the two replicas 
constitutes the "damage". Depending on the Hamiltonian, the dynamic rules, and the external
parameters the damage will either spread or heal with time (or remain in a finite spatial region). 
This behavior distinguishes chaotic or regular phases.

Kinetic Ising models are among those systems for which DS has been studied
most intensive. The majority of the work has been devoted to single-spin-flip dynamic 
rules like Glauber, Metropolis or heat-bath dynamics
\cite{derrida1,stanley,jan,costa,caer,grassberger,heat,vojta1,vojta2} but also the
Swendson-Wang cluster algorithm has been investigated \cite{swendson}. 
It has been found that the properties of DS (e.g. the question whether the damage
spreads or heals for a particular model) depend sensitively on the
dynamic rule chosen, i.e. DS is uniquely defined only if one specifies
the Hamiltonian and the dynamics. (Note that by considering all possible dynamic 
rules which are consistent with physics of a single replica an unambiguous 
definition of DS for a particular model can be obtained \cite{hinrichsen}.)

The Glauber, Metropolis or heat-bath algorithms (as well as all other single-spin-flip 
algorithms) are examples for a dynamics 
with non-conserved order parameter. There are, however, many physical systems
that can be described by kinetic Ising models with order parameter conservation.
A prominent example are, e.g., localized electrons where the
Ising variables describe the electronic occupation numbers, and the dynamics consists
of thermally assisted hops of an electron from one site to another. 
The simplest order parameter conserving dynamics in an Ising model is the spin-exchange 
dynamics of Kawasaki \cite{kawasaki}. In this paper we want to investigate DS 
for this dynamics. 
To this end we first generalize the master equation approach \cite{vojta1,vojta2} to 
dynamic rules involving more than one site. We then derive an effective-field theory
for DS in an Ising model with spin-exchange dynamics and solve it for a 
two-dimensional model on a honeycomb lattice. We find that in this model the damage 
always spreads. The stationary value of the damage is given by $D^*=(1-m^2)/2$ 
($m$ is the magnetization) which corresponds to completely uncorrelated 
configurations. The results of the effective-field theory are confirmed by Monte-Carlo
simulations.

\section{Master equation approach}
 
We consider two identical Ising models with $N$ sites described by
the Hamiltonians $H^{(1)}$ and $H^{(2)}$ given by
\begin{equation}
H^{(n)} = - \frac 1 2  \sum_{ij} J_{ij} S_i^{(n)} S_j^{(n)} 
\end{equation}
where $S_i^{(n)}$ is an Ising variable with the values $\pm 1$, and $n=1,2$ 
distinguishes the two replicas. $J_{ij}$ is the exchange interaction between the 
spins which we take to be $J$ for nearest neighbor sites and zero otherwise.
The dynamics (also called the Kawasaki dynamics \cite{kawasaki}) 
consists of exchanging spins on nearest-neighbor sites if the probability 
\begin{equation}
P=v(\Delta E/2) = \frac {e^{-\Delta E/2T}}{e^{\Delta E/2T}+e^{-\Delta E/2T}}
\label{eq:rate}
\end{equation}
is larger than a random number $\xi \in [0,1)$. Here $\Delta E$ is the energy 
change due to the exchange of the spins and $T$ denotes the temperature. 
With this dynamics the total magnetization does not change with time, i.e. 
it is a conserved quantity.

Within the master equation approach \cite{vojta1,vojta2} the 
simultaneous time evolution of the two replicas is described
by the probability distribution
\begin{equation}
P(\nu_1,\ldots,\nu_N,t) = \left \langle \sum_{\nu_i(t)} \prod_i 
\delta_{\nu_i,\nu_i(t)} \right \rangle
\end{equation}
where $\langle \cdot \rangle$ denotes the average over 
the noise realizations. The variable $\nu_i$ with the values
$++, +-, -+$,  or $--$ describes the
states of the spin pair ($S_i^{(1)}, S_i^{(2)}$). In the case of
a spin-exchange dynamics the distribution 
$P$ fulfills the master equation
\begin{eqnarray}
\lefteqn{\frac d {dt} P(\nu_1,\ldots,\nu_N,t) =} 
\label{eq:master} \\
& &- \sum_{\langle ij \rangle} \sum_{\mu_i, \mu_j}
P(\nu_1,\ldots,\nu_i,\ldots,\nu_j, \ldots, \nu_N,t) w(\nu_i,\nu_j \to \mu_i,\mu_j) \nonumber\\
& &+ \sum_{\langle ij \rangle} \sum_{\mu_i,\mu_j}
P(\nu_1,\ldots,\mu_i,\ldots,\mu_j,\ldots,\nu_N,t) w(\mu_i,\mu_j \to \nu_i,\nu_j) \nonumber
\end{eqnarray}
where $\langle ij \rangle$ denotes all pairs of nearest neighbors
and $w(\nu_i,\nu_j \to \mu_i,\mu_j)$ is the probability for a transition of the states of the sites
$i$ and $j$ from $\nu_i,\nu_j$ to $\mu_i,\mu_j$. These transition probabilities can be obtained
from (\ref{eq:rate}). In table \ref{tab:proc} we list all processes $(\nu_i,\nu_j \to \mu_i,\mu_j)$  
which lead to creation or destruction of damage, 
the probabilities for these processes will show up in the damage equation of motion
later on.
\begin{table}
\caption{Damage creating and destructing processes for spin-exchange dynamics, all
     other processes do not change the damage.}
\begin{tabular}{lcc}
&$++,-- \quad \to \quad +-,-+$ &  \\
two damaged&$++,-- \quad \to \quad -+,+-$  \\
sites created&$--,++ \quad \to \quad +-,-+$  \\
&$--,++ \quad \to \quad -+,+-$ \\
\hline\\[-2mm]
&$+-,-+ \quad \to \quad ++,--$ \\
two damaged&$+-,-+ \quad \to \quad --,++$ \\
sites destroyed&$-+,+- \quad \to \quad ++,--$ \\
&$-+,+- \quad \to \quad --,++$ \\
\end{tabular}
\label{tab:proc}
\end{table}
An important observation is that damaged sites can be created and destroyed 
only in pairs. All damage creating processes in
table \ref{tab:proc} can be transformed into each other by exchanging systems
1 and 2 and sites $i$ and $j$. Their transition probabilities are therefore also
related by symmetry. The same is true for all damage destroying processes.
Thus, it is sufficient to calculate only two independent of the probabilities
$w(\nu_i,\nu_j \to \mu_i,\mu_j)$, e.g.,
\begin{mathletters}
\label{eqs:transprob}
\begin{eqnarray}
\lefteqn{w(++,-- \quad \to \quad +-,-+) =} \\
&\qquad & \left[ v(h_i^{(2)}-h_j^{(2)}+2J) - v(h_i^{(1)}-h_j^{(1)}+2J) \right] \times \nonumber\\ 
& & \Theta(h_i^{(1)}-h_j^{(1)}-h_i^{(2)}+h_j^{(2)} ), \nonumber\\
\lefteqn{w(+-,-+ \quad \to \quad ++,--) =} \\
& & \left[ v(h_j^{(2)}-h_i^{(2)}+2J) - v(h_i^{(1)}-h_j^{(1)}+2J) \right] \times \nonumber \\
& & \Theta(h_i^{(1)}-h_j^{(1)}-h_j^{(2)}+h_i^{(2)} ) \nonumber
\end{eqnarray}
\end{mathletters}
where $h_i = \sum_j J_{ij} S_j$ is the local magnetic field of site $i$.

As in the case of Glauber or heat-bath dynamics we derive an effective-field theory by assuming
that fluctuations at different sites are statistically independent which amounts
to approximating the distribution $P(\nu_1,\ldots,\nu_N,t)$ by a product 
of single-site distributions $P_{\nu_i}(t)$. Order parameter conservation
in the two systems imposes two conditions:  $P_{++}(t)+P_{+-}(t)=const$ and 
$P_{++}(t)+P_{-+}(t)=const$.
Inserting the decomposition
\begin{equation}
P(\nu_1,\ldots,\nu_N,t) = \prod_{i=1}^N P_{\nu_i}(t)
\end{equation}
into the master equation (\ref{eq:master}) gives a system of coupled
equations of motion for the single-site distributions
\begin{eqnarray}
\frac d {dt} P_{\nu_i} = \sum_{\nu_j,\mu_i,\mu_j} [&-& P_{\nu_i} P_{\nu_j} 
W(\nu_i,\nu_j \to \mu_i,\mu_j) 
\nonumber \\
&+& P_{\mu_i}P_{\mu_j}  W(\mu_i,\mu_j \to \nu_i,\nu_j)],
\label{eq:ssmaster}
\end{eqnarray}
where $W(\nu_i,\nu_j \to \mu_i,\mu_j)$ 
is the transition probability $w$ averaged over the states $\nu$ of all sites
except for $i$ and $j$.
The total damage (Hamming distance) $D$ can be expressed in terms of the single-site distribution
$P_{\nu}$:
\begin{equation}
D = \left \langle \frac 1 {2N} \sum_{i=1}^N |S_i^{(1)} - S_i^{(2)} |
\right \rangle = P_{+-} + P_{-+}~.
\label{eq:damdef}
\end{equation}
We note, that in contrast to Ising models with a non-conserved
order parameter, the effective-field theory (\ref{eq:ssmaster}) is not very useful
in describing a {\em single} system since the only remaining
dynamic variable for a single system, viz $m$, does not change during the time evolution.
The damage is, however, not conserved and (\ref{eq:ssmaster}) constitutes
a useful mean-field theory for its time evolution.

From the single-site master equation (\ref{eq:ssmaster}) and table
\ref{tab:proc} we derive an equation of motion of the damage. 
Using some symmetry relations \cite{vojta3} 
between the transition probabilities $W$, it reads
\begin{eqnarray}
\frac d {dt} D=  -2 D^2&~&W(+-,-+ \quad \rightarrow \quad ++,--) 
\label{eq:damaster}\\
+2[(1-D)^2-m^2]&~& W(++,-- \quad \rightarrow \quad +-,-+) ~. \nonumber
\end{eqnarray}

So far the considerations have been valid for all dimensions and lattice types.
To proceed we now study a particular system, viz. a two-dimensional
Ising model on a honeycomb lattice. The same model was studied in Refs. 
\cite{vojta1,vojta2} for single-spin-flip dynamic rules.
The calculation of the average transition probabilities $W$ can be carried out
analogously to the case of single-spin-flip dynamics.
It is straightforward but tedious, the details will be published elsewhere
\cite{vojta3}. Here we only give the results:
\begin{mathletters}
\label{eqs:transition}
\begin{eqnarray}
\lefteqn{W(++,-- \quad\to\quad +-,-+)= }
\label{eq:birth}\\
&\quad&\left( - \frac{D^4} 2 + D^3 - \frac{3D^2} 4+ \frac D 2 -\frac{D^2m^2} 4+ 
\frac{Dm^2} 2  \right) t_2 +\nonumber\\
&&+\left( -\frac{D^4} {16} + \frac {D^3} 4 -\frac{3D^2} 8 + \frac D 4 +\frac {D^2m^2} 8 - 
\frac {Dm^2} 4 \right) t_4, \nonumber\\
\lefteqn{W(+-,-+ \quad\to\quad ++,--)= }
\label{eq:death}\\
&&\left( -\frac{D^4} 2 +D^3 -\frac{3D^2} 4 +\frac 1 4 -\frac{D^2m^2} 4- \frac{m^2} 4 
\right) t_2+\nonumber\\
&&+\left( -\frac{D^4} {16} +\frac 1{16} -\frac {m^2} 8 +\frac{m^4} {16} \right) t_4.\nonumber
\end{eqnarray}
\end{mathletters}
Here $t_n=\tanh(nJ/T)$.
If we insert (\ref{eq:death}) into the damage equation of motion (\ref{eq:damaster})
we observe that the death term [first line of eq. (\ref{eq:damaster})] is of order  
$D^2$ for small $D$. This is a major difference to the case of Glauber and heat bath dynamics 
\cite{vojta1,vojta2} (where the death term is of order $D$) and reflects the fact 
that for spin-exchange dynamics the damage can only be destroyed pairwise.
In contrast, the birth term is of order $D$ (as it is for Glauber oder heat bath 
dynamics) because already a single damaged site can produce further damage
in its neighborhood. Consequently, for small enough $D$ the birth term will always be
larger than the death term and the damage will never heal completely.

\begin{figure}
  \epsfxsize=\figuresize
  \centerline{\epsffile{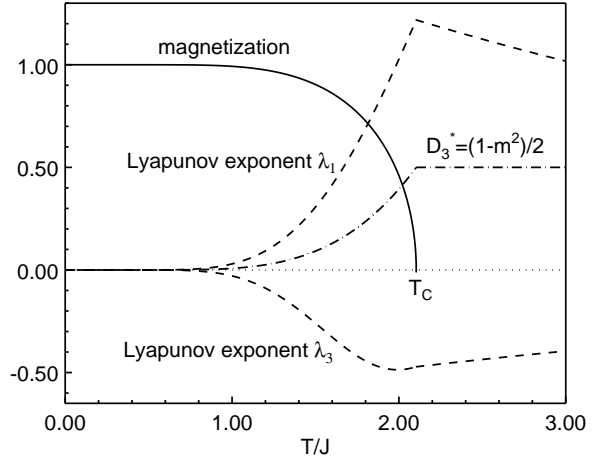}}
  \caption{Results of the effective-field theory for damage spreading
              in the kinetic Ising model with spin exchange dynamics.}
  \label{fig:lia_kawa}
\end{figure}

We now discuss the stationary solutions of the damage equation of motion 
(\ref{eq:damaster}) and their stability. We restrict ourselves to the case that
the system
is in equilibrium when the damage is introduced. Thus, $m$ can be taken to 
be the equilibrium value of the magnetization which is zero for $T>T_c \approx 2.11 J$ and 
\begin{equation}
m^2 = \frac { \frac 3 4 (\tanh 3J/T + \tanh J/T) -1}  { \frac 3 4 \tanh J/T 
- \frac 1 4 \tanh 3J/T}
\end{equation}
for $T<T_c $ in our effective-field theory \cite{vojta1}.
Obviously, $D=D_1^*=0$ is always a fixed point (FP) of (\ref{eq:damaster}).
To investigate its stability we expand (\ref{eq:damaster}) to linear order in $D$.
The resulting linearized equation of motion is given by $dD/dt = \lambda_1 D$
with 
\begin{equation}
\lambda_1 = (1-m^4) \tanh\frac {2J} T + \frac 1 2 (1-m^2)^2 \tanh\frac {4J} T  .
\label{eq:lambda_1}
\end{equation}
The Lyapunov exponent $\lambda_1$ is always positive, thus the FP
$D_1^*$ is always unstable. In Fig. \ref{fig:lia_kawa} we show the
temperature dependence of $\lambda_1$.
Since the Hamiltonian and the dynamic rule are invariant under a global flip of all spins,
the existence of the FP $D_1^*=0$ implies the existence of 
the FP $D_2^*=1$ with the same stability properties.

In the paramagnetic phase the only other stationary solution of
(\ref{eq:damaster}) in the physical interval $0\le D \le1$ is $D_3^*=1/2$. 
By expanding (\ref{eq:damaster})
around $D=1/2$ we obtain the corresponding Lyapunov exponent,
\begin{equation}
\lambda_3 = - \frac 3 8 \tanh \frac {2J} T - \frac {13} {64} \tanh \frac {4J} T.
\label{eq:lambda_3}
\end{equation}
Since $\lambda_3$ is always negative the FP $D_3^*=1/2$ is 
stable in the entire paramagnetic phase. The temperature dependence 
of $\lambda_3$ is also shown in Fig. \ref{fig:lia_kawa}.

In the ferromagnetic phase there are two more stationary solutions
of (\ref{eq:damaster}) in addition to $D_1^*=0$ and $D_2^*=1$. 
If the two systems have the same value of the magnetization,
$m^{(1)}=m^{(2)}=m$ we obtain the FP 
$D_3^*=(1-m^2)/2$. If the two systems have opposite magnetization
$m^{(1)}=-m^{(2)}=m$ we obtain the FP $D_4^*=(1+m^2)/2$.
Since $D_3^*$ and $D_4^*$ are related by a global flip of all spins
in one of the systems, they have the same stability properties.
The corresponding Lyapunov exponents are given by
\begin{eqnarray}
\lambda_3 = \lambda_4 &=& \left( -\frac 3 8 - \frac {m^2} 2 + 
     \frac {m^4} 4 + \frac {m^6} 2 + \frac {m^8} 8 \right) \tanh \frac {2J} T \\
\label{eq:lambda_4}
&+&\left(-\frac {13} {64} + \frac {5m^2} {16} + \frac {m^4} {32} -
     \frac {3m^6} {16} + \frac {3m^8} {64} \right) \tanh\frac  {4J} T \nonumber.
\end{eqnarray}
They are always smaller than zero, thus $D_3^*$ and $D_4^*$ are stable
in the entire ferromagnetic phase. The temperature dependence of 
$D_3^*$ and $\lambda_3$ is shown in Fig. \ref{fig:lia_kawa}.

We now show that at the stable FPs the configurations of the
two systems are completely uncorrelated. From the definition 
(\ref{eq:damdef}) of the damage we obtain
\begin{equation}
D = \frac 1 {2N} \sum_{i=1}^N  \left \langle |S_i^{(1)} - S_i^{(2)} | \right \rangle 
   = \frac 1 {2N} \sum_{i=1}^N (1- \langle S_i^{(1)}  S_i^{(2)}   \rangle ).
\end{equation}
For uncorrelated configurations of $S_i^{(1)}$ and $S_i^{(2)}$ we have 
$\langle S_i^{(1)}  S_i^{(2)}  \rangle=\langle S_i^{(1)} \rangle\langle S_i^{(2)} \rangle=\pm m^2$
if the two systems have equal or opposite magnetization, respectively.
Thus, for uncorrelated configurations we obtain $D=(1 \mp m^2)/2$. These are exactly
the values of $D_3^*$ and $D_4^*$. 

\section{Monte-Carlo simulations}

We have verified the main predictions of the mean-field theory by Monte-Carlo
(MC) simulations of a three-dimensional Ising model with Kawasaki spin-exchange 
dynamics according to (\ref{eq:rate}). The simulations are carried out on cubic lattices  
with up to $N=101^3$ sites with periodic boundary conditions. By comparing
different system sizes we verify that any finite size corrections to the results
are smaller than the statistical error of the simulation. This is easily possible
since we are away from a spreading transition and thus the damage correlation 
length is finite.

\begin{figure}
  \epsfxsize=\figuresize
  \centerline{\epsffile{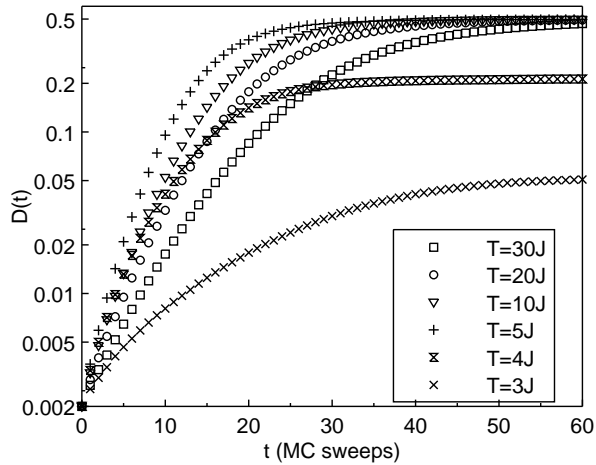}}
  \caption{Time evolution of the damage in an Ising model with spin-exchange 
               dynamics. The data points represent averages over 100 runs of a system
               of $27^3$ sites.}
  \label{fig:evo}
\end{figure}

In this study we are not interested in phase separation processes. We thus prepare
the system with the correct equilibrium magnetization value for each temperature
simulated. 2000 MC sweeps are carried out to equilibrate the system.
Then the initial damage $D_0$ is created by exchanging
randomly chosen pairs of nearest neighbor spins in one of the systems.
We use values of $D_0$ between $5*10^{-4}$ and $5*10^{-2}$.
After that both systems evolve in parallel using the same random numbers.
Examples of the time evolution of the damage are shown in Fig. \ref{fig:evo}.
Within the first 5 to 10 MC sweeps the damage increases approximately
exponentially with time. A fit of the data to an exponential law gives an estimate 
for the Lyapunov exponent $\lambda_1$. We note that the damage time evolution
shows a systematic deviation from an exponential law which manifests in a slight
downward curvature in Fig. \ref{fig:evo}. This deviation stems from the fact
that we are simulating a lattice system. Since with Kawasaki dynamics the damage can spread 
at most two lattice constants per time step the increase of the damage with time is bounded
by a power law,  $D(t) \le D_{max} \sim (2t)^3 D_0$. Therefore, a pure exponential spreading 
can only be observed
as long as the probability for any site (or pair of sites) to become damaged 
during a particular time step is small compared
to one. (In this case the above bound set by the lattice does not play a role.)
For our system this condition is, however, only fulfilled for small temperatures.

\begin{figure}
  \epsfxsize=\figuresize
  \centerline{\epsffile{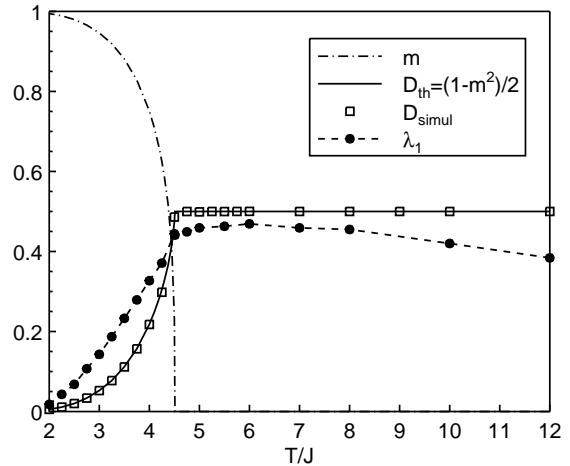}}
  \caption{Asymptotic average damage $D_3^*$ and Lyapunov exponent $\lambda_1$
             for the kinetic Ising model with spin exchange dynamics. The Lyapunov 
             exponents have been obtained from 100 runs of a $27^3$ system. The FP 
             values $D_3^*$ has been calculated from 10 runs of a $101^3$ system.
             Their statistical errors are smaller than the symbol size.}
  \label{fig:simu}
\end{figure}

In order to determine the long-time limit of the average damage we average its values 
over 5000 MC sweeps after a plateau has been reached.
The results of our simulations are summarized in
Fig. \ref{fig:simu}. We indeed find that the FP $D_1^*$ is unstable, and the damage 
always spreads. The Lyapunov exponent $\lambda_1$ of the
FP $D_1^*$ is positive for all temperatures investigated.  The
asymptotic average damage takes exactly the value of two uncorrelated
configurations, viz. $D_3^*=(1-m^2)/2$. (We did not observe the other stable 
FP $D_4^*=(1+m^2)/2$ since we always started with the two systems having the 
same magnetization.)

\section{Conclusions}

To summarize, we have used an effective-field theory and Monte-Carlo
simulations to show that the time evolution of a kinetic Ising model
with  Kawasaki spin exchange dynamics is chaotic for all temperatures
in the sense that the FP $D_1^*=0$ is unstable. 
Moreover, we have shown that two systems whose initial configurations 
differ only at a few sites become completely uncorrelated in the long-time limit.
This corresponds to an asymptotic average damage of $D=(1-m^2)/2$. 

\begin{figure}
  \epsfxsize=\figuresize
  \centerline{\epsffile{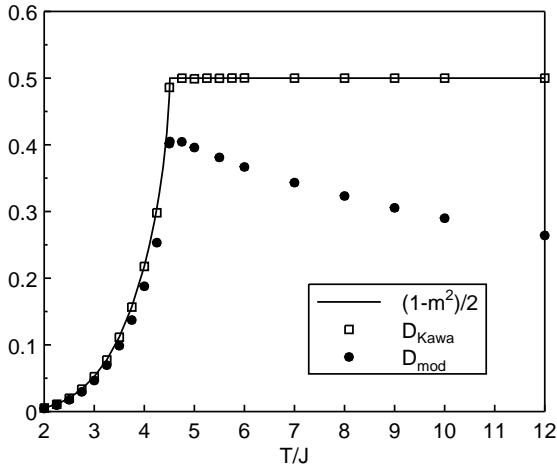}}
  \caption{Comparison of the asymptotic average damage $D_3^*$ 
             for the kinetic Ising model with Kawasaki and modified spin exchange dynamics.}
  \label{fig:mod}
\end{figure}

In this last part of the paper we want to discuss how general these results are.
Since the properties of DS are known to depend on how the random numbers are used 
in the update process \cite{hinrichsen} for single-spin-flip dynamics,
an analogous comparison for spin-exchange dynamics is desirable.
However, the main properties of our solution will be robust against
such changes in the update rules. In particular, the fact that the damage 
death rate (see eq. \ref{eq:damaster}) is of order $D^2$ is a result of the
spin-exchange mechanism alone. It is therefore independent of how the random numbers 
are used in the update rule. This suggests that the main finding
of this paper, viz. that the spin-exchange dynamics is chaotic for all temperatures
is valid not only for the Kawasaki update rule (\ref{eq:rate}) but in general.
As a first step of a future systematic investigation of different update rules we
have studied a modified version of the Kawasaki dynamics.  The modification
consists of using the random number $\xi$ if the configuration of the spin 
pair selected for the exchange is $(+-)$ but using $1-\xi$ instead if the configuration 
is $(-+)$. This modified update rule can be seen as the spin-exchange analog
of the heat-bath dynamics (in the same sense as the Kawasaki dynamics can be seen 
as the analog of the Glauber dynamics). 
In Fig. \ref{fig:mod} we compare the asymptotic average damage of the Kawasaki
and the modified spin-exchange dynamics.
The modified dynamics gives lower damage values than the Kawasaki dynamics for all
temperatures. Nonetheless, the fixed point $D_1^*=0$ is unstable for all finite temperatures,
and the asymptotic damage is finite. This is in agreement with the above suggestion that
a spin-exchange dynamics is always chaotic irrespective of the particular update rule.

Let us finally discuss the relation of the DS process discussed here with other 
non-equilibrium processes. As already mentioned, a key feature of  DS with spin-exchange
dynamics is that damaged sites can heal only in pairs while they can diffuse alone
and also create further damage. This is different from the contact process and other
processes in the directed percolation universality class where a single active site can die
locally with finite probability. There is, however, a simple reaction-diffusion process
which should show qualitatively the same behavior as DS with spin-exchange dynamics.
Since for small damage $D$ the birth rate is proportional to $D$  (and since damage is created
in pairs) while the death rate is
proportional to $D^2$ (see eqs. (\ref{eq:damaster},\ref{eqs:transition}), such a reaction-diffusion  
process could be defined by the reactions 
\begin{eqnarray}
  A &~&\stackrel{p_1}{\longrightarrow} ~3A\\
 2A &~&\stackrel{p_2}{\longrightarrow} ~0 \nonumber
\end{eqnarray}
and additional diffusion of the substance A. For small concentrations of A it
should have the same qualitative behavior as DS with spin-exchange dynamics for small
damage.

This work was supported in part by the DFG under grant
numbers Vo659/1-1 and SFB393 and by the NSF under grant number 
DMR-95-10185.


\end{document}